\documentclass[twocolumn,showpacs,preprintnumbers,amsmath,amssymb]{revtex4}

\newcommand{\ds}{\displaystyle}

\newcommand{\be}{\begin{equation}}
\newcommand{\en}{\end{equation}}
\newcommand{\bea}{\begin{eqnarray}}
\newcommand{\ena}{\end{eqnarray}}

\usepackage{graphicx}% Include figure files
\usepackage{bm}% bold math

\begin{document}

\title{Extended open inflationary universes}

\author{ Sergio del Campo
  and Ram\'on Herrera }
\affiliation{ Instituto de F\'\i sica, Universidad Cat\'olica de
Valpara\'\i so, Av. Brasil 2950, Casilla 4059, Valpara\'\i so,
Chile.}

\date{\today}% It is always \today, today,

\begin{abstract}
In this paper we study a type of one-field model for open
inflationary universe models in the context of the
Jordan-Brans-Dicke theory. In the scenario of a one-bubble
universe model we determine and characterize the existence of the
Coleman-De Lucia instanton, together with the period of inflation
after tunnelling has occurred. Our results are analogous to those
found in the Einstein General Relativity models.
\end{abstract}

\maketitle

\section{\label{sec:level1} Introduction}

Until recently, inflation~\cite{AG,AAPS,AL} was always associated
with a flat universe, due to its ability to drive the spatial
curvature  to zero so effectively. In fact, requiring sufficient
inflation to homogenize random initial conditions drives the
universe very close to critical density. However, current
observations, e.g. the cosmic microwave background (CMB)
radiation, show a large degree of homogeneity, but are as yet
inconclusive as to spatial curvature. Because of this, some
authors have put forward the idea of considering other models in
which the curvature does not vanished.

In the context of an open scenario, it is assumed that the
universe has a lower-than-critical matter density and, therefore,
a negative spatial curvature. Several
authors~\cite{re3,re4,re5,re6}, following previous speculative
ideas~\cite{re1,re2}, have proposed possible scenarios in which
open universes may be realized, and its consequences, such that
density perturbations, have been explored~\cite{UMRS}. Very
recently the possibility  has been considered  to create an open
universe from the perspective of the brane-world
scenarios~\cite{MBPGD}.

The need to consider open models is reflected in the fact that
primordial perturbations and their corresponding power spectrum
give rise in a natural way not only to adiabatic perturbation but
also to isocurvature perturbations. This generic situation is
obtained, for instance, in the case  where two fields are excited
during inflation. We expect this sort of situation in a model in
which the inflaton $\sigma$ and the Jordan-Brans-Dicke (JBD)
$\phi$ scalar fields are present~\cite{KEHKSJV}.

The basic idea in an open universe is that a symmetric bubble
nucleates in the de Sitter space background, and its interior
undergoes a stage of slow-roll inflation, where the parameter
$\Omega_0$ can be adjusted to any value in the range $0 < \Omega
_0 < 1$.

Bubble formation in the false vacuum is described by the
Coleman-De Lucia(CDL) instantons~\cite{re7}. Once a bubble has
taken place by this mechanism, the bubble's inside looks like an
infinite open universe. The problem with this sort of scenario is
that the instanton exists only if the following inequality $\mid
V^{''}\mid > H^{2}$ is satisfied during the tunneling process. On
the contrary, during inflation the inequality $\mid V^{''}\mid \ll
H^{2}$ is satisfied (slow-roll approximation).  From now on
$V^{''}$ stands for $d^{2}V/d\sigma^{2}$, where $\sigma$ is the
inflaton scalar field and $V = V(\sigma)$, the inflaton potential.
Linde solves this problem by proposing a simple one-field model in
Einstein's general relativity (GR) theory~\cite{re6}(see also
ref.~\cite{re8}). At this point, we should mention that Ratra and
Peebles were the first to elaborate on the open inflation
model~\cite{RaPe}.

In ref.~\cite{re9}  a one-field model for open inflation by using
scalar-tensor type of theory (a nonminimally coupled scalar field
with  polynomial potentials)is studied. Here, the scalar potential
associated with the JBD field is assumed to be $\ds V(\phi)
=\frac{1}{2}m^2 \phi^2 (1+f^2(\phi))$. Then, the arbitrary
function $f(\phi)$ is fixed in such a way that in the Einstein
frame the resulting effective action coincides with the one used
by Linde~\cite{re6}. This certainly restricts not only the
evolution of the JBD field but also the parameters that enter into
the model.

The purpose of the present paper is to study a one-field open
inflation in a JBD theory~\cite{re10}, where the inflaton field
$\sigma$ is of the same nature as that described by
Linde~\cite{re6}. In this sense, our model will be a genuine,
extended open inflationary universe model.

The plan of the paper is as follows: In Sec. II we numerically
write and solve  the field equations in a Euclidean spacetime.
Here, the existence of the CDL instanton for two different models
are described. In Sec. III we determine the characteristic of the
open inflationary universe model that is produced after tunnelling
has occurred. In Sec. IV we determine the corresponding density
perturbations for our models. Our results are compared with the
analogous results obtained by using the Einstein theory of
gravity. Finally, we conclude in Sec. V.

%%%%%%%%%%%%%%%%%%%%%%%%%%%%%%%%%%%%%%%%%%%%%%%%%
%                                               %
%           ECUACIONES DE CAMPO  EUCLIDEANAS    %
%                                               %
%%%%%%%%%%%%%%%%%%%%%%%%%%%%%%%%%%%%%%%%%%%%%%%%%

\section{\label{sec:level2}The Euclidean cosmological equations in JBD theory }

We consider the effective action given by
 \be \hspace{0.cm} \ds
S\,=\,\int{d^{4}x\,\sqrt{-g}}\,\left
[\,\frac{1}{2}\,\varepsilon\,\phi^{2}\,R\,
-\,\frac{1}{2}\,\partial_{\mu}\phi \,\partial^{\mu}\phi
-{\cal{L}}(\sigma)\right], \label{ac1}
 \en
where
$${\cal{L}}(\sigma)\,=\,\frac{1}{2}\partial_{\mu}\sigma
\partial^{\mu}\sigma\,+\, V(\sigma),$$
%\right ], \label{ac1}
%\end {equation}
and $R$ is the Ricci scalar curvature, $\phi$ is the JBD scalar
field, and $ \varepsilon$ is a dimensionless coupling constant
that, in terms of JBD parameter $\omega$, is equivalent to
$1/4\,\varepsilon $. $V(\sigma)$is an effective scalar potential
associated with the inflaton field, $ \sigma$.

The $O(4)$- invariant Euclidean spacetime metric is described as
\be \ds d{s}^{2}\,=\, d{\tau}^{2}\,+\, a(\tau)^{2}\, (\,\,
d{\psi}^{2}\,+\,\sin^{2}\psi\,d{\Omega^{2}_{2}}\,\,), \label{met}
\en where $a(\tau)$ is the scale factor, and $\tau$ represents the
Euclidean time.

When metric~(\ref{met}) is introduced into action~(\ref{ac1}), we
obtain the following field equations: \be \ds
\left(\frac{a^{,}}{a}\right)^{2}+2
\frac{a^{,}}{a}\frac{\phi^{,}}{\phi}=\frac{1}{a^{2}}+
\frac{1}{3\varepsilon\phi^{2}}\left[\frac{\phi^{,\,
2}}{2}+\frac{\sigma^{,\,2}}{2}-V(\sigma) \right]\, \label{ec2},
 \en
\be \ds \phi^{,,}+3\frac{a^{,}}{a}\phi^{,}+\frac{\phi^{,\,2}}{\phi}
+\frac{1}{1+6\varepsilon}\left[\frac{\sigma^{,\,2}}{\phi}+\frac{4}{\phi}V(\sigma)
\right]=0\,\, \label{ec3},
 \en
\be \ds
\sigma^{,,}\,=\,-3\,\frac{a^{,}}{a}\,\sigma^{,}\,+\,\frac{dV}{d\sigma}\,\,\label{ec4},
\en where the primes denote derivatives with respect to $\tau$.
 From now on we will use units where c =
$\hslash$=M$_{p}$=G$^{-1/2}$ = 1.

 From Equations~(\ref{ec2}),~(\ref{ec3}) and~(\ref{ec4}) we obtain
\be \ds
a^{,,}=2\frac{a^{,}\phi^{,}}{a\phi}-\frac{a}{3\varepsilon\phi^{2}}\left[
\phi^{,2}+\frac{1+3\varepsilon}{1+6\varepsilon}\sigma^{,\,2}+
\frac{1-6\varepsilon}{1+6\varepsilon}V(\sigma)\right]. \label{ec5}
\en

The first model  considered corresponds to the effective potential
used by Linde\cite{re6}: \be \ds
V(\sigma)\,=\,\frac{m^{2}\,\sigma^{2}}{2}\left(\,1\,+
\,\frac{\alpha^{2}}{\beta^{2}+(\sigma-v)^{2}}\right)\,,\label{pot1}
 \en
where $\alpha$, $\beta$ and $v$ are arbitrary constants. In this
potential the first term controls inflation after quantum
tunneling has occurred. Its form coincides with that used in the
simplest chaotic inflationary universe model, $m^{2}\sigma{^2}/2$.
The second term controls the bubble nucleation, whose role is to
create an appropriate shape in the inflaton potential,
$V(\sigma)$, where its maximum occurs near $\sigma$ = $v$.
Following Linde\cite{re6} we take $\beta^{2}$=2$\alpha^{2}$,
$\beta$=0.1 $v$=3.5 and $m$= 1.5 $\times$10$^{-6}$. Certainly,
this is not the only choice, since other values for these
parameters can also lead to a successful open inflation scenario
(with any value of $\Omega$, from 0 and 1).

\begin{figure}[ht]
\includegraphics[width=3.0in,angle=0,clip=true]{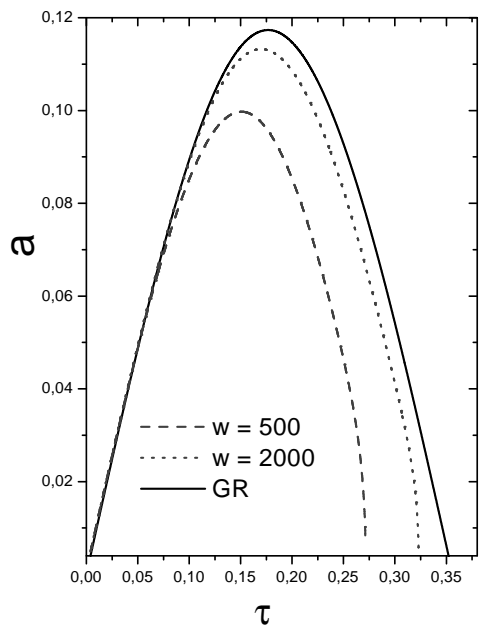}
\caption{For our model we plot the scale factor a($\tau$) as a
function of the Euclidean time $\tau$, for two different values of
the JBD parameter, $\omega$=500 and $\omega$= 2000. GR represents
the same graph, but it was obtained by using the Einstein GR
theory. We have assumed constant $m$ to be equal to one.}
\label{fe}
\end{figure}

\begin{figure}[ht]
\includegraphics[width=3.0in,angle=0,clip=true]{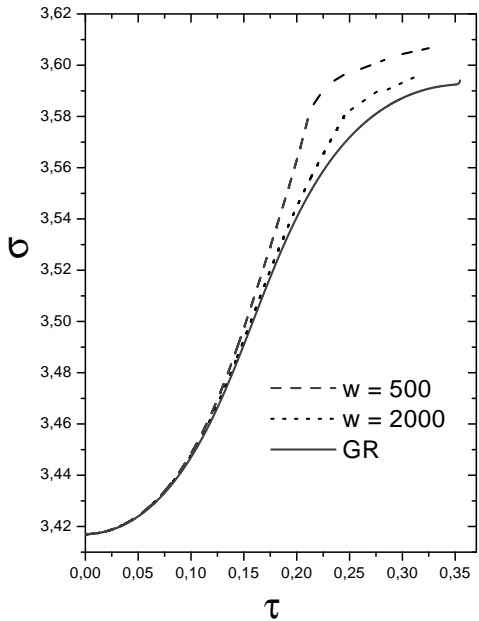}
\caption{The instanton $\sigma(\tau)$ as a function of the
Euclidean time $\tau$ is shown for Einstein's GR and JBD theories.
In the latter case we have assumed $\omega = 500$ and $\omega =
2000$. As before, we have assumed $m = 1$.} \label{ane}
\end{figure}

We have solved the field Equations~(\ref{ec2})-(\ref{ec4})
numerically. The boundary conditions that we used are those in
which $\phi^{,}$=$\sigma^{,}$=0 and $a^{,}= 1$ at $a=0$, for
various values of the JBD parameter, $\omega$. At $\tau \approx
0$, the scalar field $\sigma=\sigma_T$ lies in the "true vacuum",
near the maximum of the potential, $V(\sigma)$, and at $\tau \neq
0$, the same field is found closed to the false vacuum, but now
with a different value, $\sigma = \sigma_F$. In our model, when
the scalar field $\sigma$ evolves from some initial value
$\sigma_{F}\cong\sigma_{i}\approx 3.6$ to the final value
$\sigma_{T} \cong \sigma_{f}\approx 3.4$, we found that the CDL
instanton does exist, and the extended open inflationary universe
scenario can be realized. Figure~\ref{fe} shows how the scale
factor evolves during the tunneling process.
Note that the interval of tunneling, specified by $\tau$,
decreases when the parameter $\omega$ decreases, but its shapes
remain practically similar. The evolution of the inflaton field as
a function of the Euclidean time is shown in Fig.~\ref{ane}. Note
the similar quantities that $\sigma(\tau)$ contracts at the
beginning of the inflationary era ($\tau = 0$).

\begin{figure}[ht]
\includegraphics[width=3.0in,angle=0,clip=true]{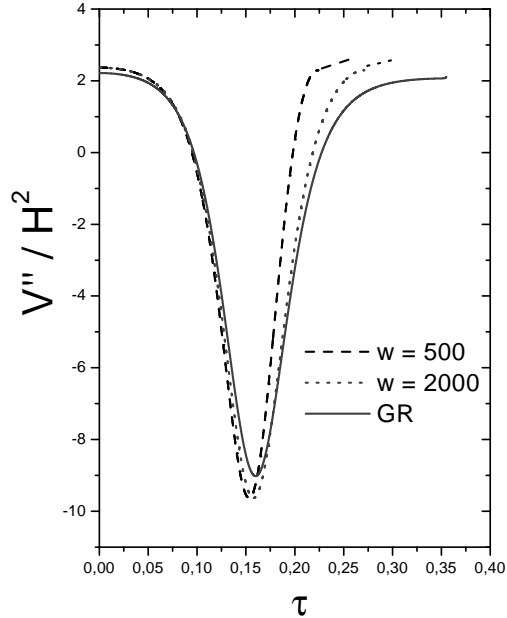}
\caption{This plot shows how during the tunneling  process the
inequality $|V^{''}|\gg H^{2}$ is satisfied for two values of
$\omega$($\omega$=500 and $\omega$=2000 for JBD theory). The
continuous line represents the same inequality for Einstein's GR
theory. Again, we have taken  $m=1$. } \label{fig1}
\end{figure}

In Fig.~\ref{fig1} we show  $|V^{''}|\,/\,H^{2}$ as a function of
the Euclidean time $\tau$ for our model. From this plot we observe
that, most of the time during the tunneling, we obtain
$|V^{''}|\,>\,H^{2}$, analogous to what occurs in the Einstein's
GR theory. Note that, as far as we decrease the value of the
parameter $\omega$, the peak becomes narrower and deeper, and thus
the above inequality is better satisfied.
In our model, it is possible to numerically show that the CDL
instanton $\sigma(\tau)$ exists, and for various values of the
$\omega$ parameter, it presents a similar behavior to that
described in the Linde's paper~\cite{re6}. The values actually
coincide for small $\tau$, and its values (after tunneling has
occurred) coincide in the two theories, i.e. Einstein's GR and JBD
theories. This result shows that the value that $\sigma$ obtains
at the end of the tunneling process is independent of the $\omega$
parameter. On the other hand, the numerical solution shows that
the evolution of the JBD field $\phi$ during the tunneling process
is such that  it remains practically constant for $\tau \leq 0.3$,
and it then decreases for $\tau > 0.3$.

\begin{figure}[ht]
\includegraphics[width=3.0in,angle=0,clip=true]{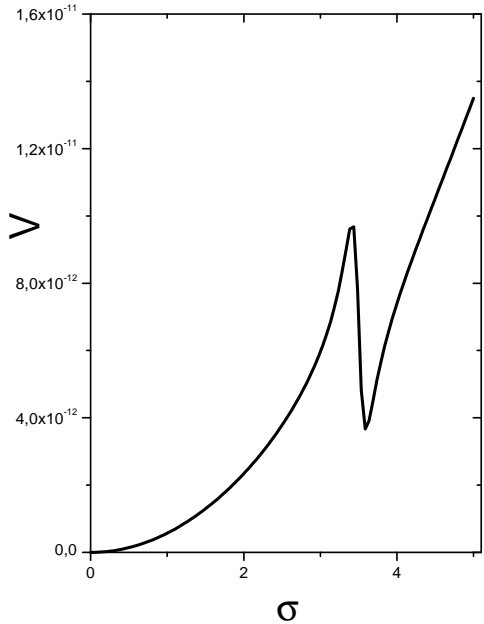}
\caption{The effective potential for our second model. Note the
similarity between this potential and that considered in
Ref.~\cite{re8}. } \label{potencial}
\end{figure}

At this point, we would like to consider a new effective
potential: \be \ds
V(\sigma)\,=\,\frac{m^{2}\,\sigma^{2}}{2}\left(\,1\,+
\,\frac{\alpha^{2}\,tanh(v-\sigma)}{\beta^{2}+(\sigma-v)^{2}}\right)\,.\label{pot2}
\en which is quite similar to that studied in
Reference~\cite{re8}. Here $\alpha$, $\beta$ and $v$ are arbitrary
constants. For completeness, we will restrict ourselves to the
particular case in which the different constants take the values m
= 1.5 $\times$ 10$^{-6}$, $v$ = 3.5, $\alpha^{2}$ = 0.1, and
$\beta^{2}$ = 0.01. The shape of this potential is shown in
Fig.~\ref{potencial}. We should mention that both effective
potentials~(\ref{pot1}) and~(\ref{pot2}) present a pronounced
peak, which is necessary for open inflation to occur. The nature
of these potentials may be varied. For instance, present-day
supersymmetry and supergravity theories include many scalar
fields, whose interaction potentials may be arbitrary to certain
extent. It is therefore useful to study the de Sitter stage which
is produced by effective potentials, such as those expressed by
Eqs.~(\ref{pot1}) and~(\ref{pot2}). As Linde mentions in
Ref.~\cite{re8}, when  $\sigma \longrightarrow v$ a sharp peak
appear in the effective potential, and it may be due to the
emergence of a strong coupling regime in the Yang-Mills sector,
where the energy density gets a contribution from new terms into
the Lagrangian, such that $<F_{\mu \nu}^a\,F^{\mu \nu}_a>$, where
$F_{\mu \nu}^a =
\partial_\mu A^a_\nu - \partial_\nu A^a_\mu + e T^a_{b c}A^b_\mu
A^c_\nu $ are the field strengths for the Yang-Mills fields
$A_\mu^a$, where $e$ is the gauge coupling constant, and the
$T^a_{b c}$ are the structure constants of the Lie algebra.

\begin{figure}[ht]
\includegraphics[width=3.0in,angle=0,clip=true]{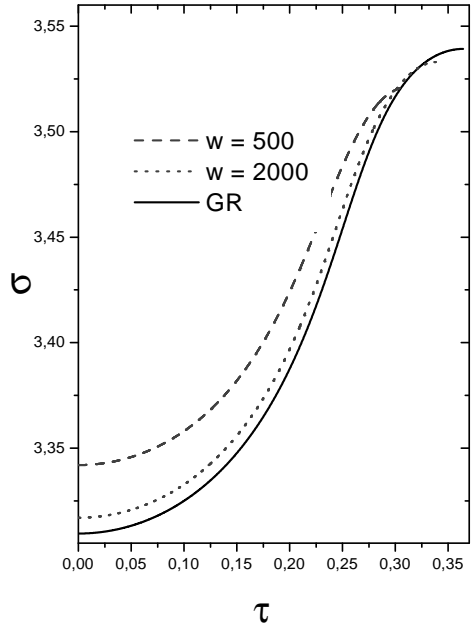}
\caption{The instanton $\sigma(\tau)$ as a function of the
Euclidean time $\tau$ for our second model in both Einstein's GR
and JBD theories (for $\omega = 500$ and $\omega = 2000$ in the
latter case). Note the different values that occur for $\sigma$ at
the beginning of inflation, i.e. at $\tau = 0$. Again, as before,
we have assumed $m = 1$.} \label{sigma2}
\end{figure}

The Coleman-De Luccia instanton $\sigma(\tau)$ in this model is
shown in Fig.~\ref{sigma2}. Our results are compared to that
corresponding to the Einstein's GR theory. Tunneling occurs from
the initial point $\sigma_{i}\approx 3.54$, which almost coincides
with the local minimum of V($\sigma$), to final point
$\sigma_{f}\approx 3.31$. The evolution of the inflaton field
during the tunneling process shown by Fig.~\ref{sigma2} is quite
similar to what happens in the previous case, but the values that
the inflaton field gets immediately after the tunneling are
different. The reason for this is due to the fact that we have
consider $d\sigma/d\tau = 0$, both at the beginning and at the end
of the tunneling. Thus, at unlike of the first case, and in order
to satisfy this condition in our second model, we were forced to
consider different initial values of the inflaton field, when
different values of the JBD parameter $\omega$ were taken.

\begin{figure}[ht]
\includegraphics[width=3.0in,angle=0,clip=true]{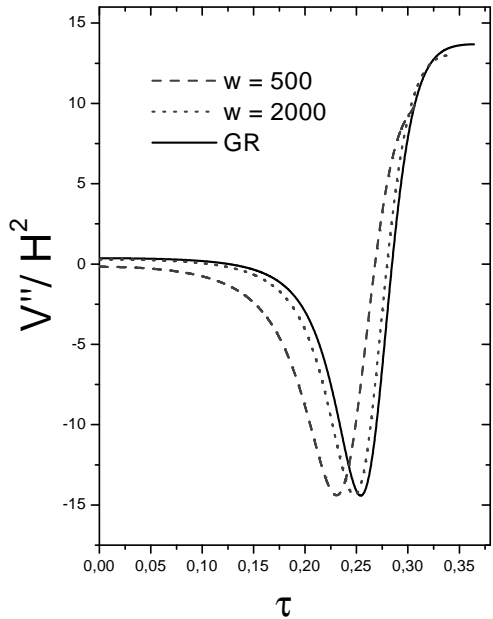}
\caption{This graph shows how the inequality
$\mid\,V''\,\mid\gg\,H^{2}$ is satisfied during the tunneling
process. The continuous line represents what happen in Einstein's
GR theory. The other curves correspond to $\omega$=500 and
$\omega$=2000 in JBD theory. We have assumed $m = 1$.}
\label{deriv2}
\end{figure}

Figure~\ref{deriv2} shows that almost everywhere along the
evolution of the scalar inflaton field, $\sigma(\tau)$, it is
found that $\mid\,V''\mid\,\gg\,H^{2}$. Unlike the previous case,
the width of the peak increases when the $\omega$ parameter
decreases.

In the following we are going to calculate the instanton action
for the quantum tunneling between the false and the true vacuum in
the JBD theory. By integrating by parts and using the Euclidean
equations of motion, we find that the action may be written as
 \be \ds
 S\,=\,2\pi^{2}\,\int\,d\tau[a^{3}(\phi^{,\,2}+\sigma^{,\,2})-
 6\varepsilon\phi\,aa^{,}(a^{,}\phi+2a\phi^{,})]. \label{action1}
\en Note that this action coincides with that corresponding to its
analogous in the Einstein's general relativity theory if we assume
that $\epsilon\phi^{2}= Cte.=1/8\pi=1/\kappa^{2}$ ~\cite{Parker}.

The $\sigma$ inflaton field is initially trapped in its false
vacuum, whose value is $\sigma_{F}$, and where the JBD field has
the value $\phi_{F}$. After tunneling to the true vacuum, the
instanton and the JBD fields  get the values $\sigma_{T}$ and
$\phi_{T}$, respectively, and a single bubble is produced .
Similar to the case of GR theory, the instanton  (or bounce)
action is given by $B = S - S_{F}$, i.e. the difference between
the action associated with the bounce solution and the false
vacuum. This action determines the probability of tunneling for
the process. We have defined $V_{F}=V(\sigma_{F})$ and
$V_{T}=V(\sigma_{T})$ as the false and true vacuum energies,
respectively. Under the approximation that the bubble wall is
infinitesimally thin, we obtain the reduced action for the
thin-wall bubble:
$$
\hspace{-1.8cm}
S=2\pi^{2}S_{1}R^{3}-4\pi^{2}[(1-H_{F}^{2}R^{2})^{3/2}-1]
\frac{\varepsilon\phi_{F}^{2}}{H_{F}^{2}}
$$
 \be \ds \hspace{3.2cm}
+\,4\pi^{2} [(1-H_{T}^{2}R^{2})^{3/2}-1]
\frac{\varepsilon\phi_{T}^{2}}{H_{T}^{2}}, \label{a10} \en where
we have taken into account the contributions from the wall (first
term) and the interior of the bubble (the second and third terms).
Here $R$ is the radius of the bubble,
$H_{F}^{2}=V_{F}/3\varepsilon\phi_{F}^{2}$ and
$H_{T}^{2}=V_{T}/3\varepsilon\phi_{T}^{2}$. The surface tension of
the wall becomes defined by
 \be \ds
 S_{1}\,=\,\int\,d\tau\,[\,\phi_{wall}^{,\,2}\,+\,\sigma^{,\,2}\,] \label{acc2},
 \en
or equivalently
 $$
S_{1}\,=\,\int\,d\tau\,\phi_{wall}^{,\,2}\,+
\,\int_{\sigma_{T}}^{\sigma_{F}}d\sigma[2(V(\sigma)-V_{F})]^{1/2}
,
$$
where $\phi_{wall}^{,}$ is the variation of scalar field $\phi$
across the bubble wall. To continue, we have taken the approach
followed by the authors of Ref.~\cite{Accetta}, where they use the
approximation $\phi_{wall}^{,} \approx 0$; in this way, we could
drop the first term of Eq.~(\ref{acc2}). However, we should note
that in our case we are concerned with the decay of a false vacuum
with positive energy density to a true vacuum in which  this
energy is also positive, but smaller than the other one, i. e.,
the decay from $V(\sigma_F)$ to $V(\sigma_{T}) \neq 0$.

The curvature radius of the bubble wall is one for which the
bounce action~(\ref{a10}) is an extremum. Then, the wall radius is
determined by setting $dS/dR$ = 0, which gives
$$
 \frac{S_{1}R}{2}=(1-H_{T}^{2}R^{2})^{1/2}\varepsilon\phi_{T}^{2}-
 (1-H_{F}^{2}R^{2})^{1/2}\varepsilon\phi_{F}^{2}.
$$
This can be solved for the radius of the bubble, and it is found that
 \be \ds
 R=\left[\frac{S_{2}\pm(
 S_{2}^{2}-H_{2}[\,1-(\phi_{T}/\phi_{F})^{4}\,]\,)^{1/2}}{H_{2}}\right]^{1/2},\label{r}
 \en
where $S_2$ is given by
$$
 S_{2}=\left(\frac{S_{1}}{2\varepsilon\phi_{F}^{2}}\right)^{2}\left[
 1+\left(\frac{\phi_{T}}{\phi_{F}} \right)^{4} \right]-\left[
 1-\left(\frac{\phi_{T}}{\phi_{F}} \right)^{4} \right][H_{T}^{2}-H_{F}^{2}],
$$
and
$$
\hspace{-1.8cm}
H_{2}=\left[H_{F}^{2}-H_{T}^{2}\left(\frac{\phi_{T}}{\phi_{F}}
\right)^{4}+\left(\frac{S_{1}}{2\varepsilon\phi_{F}^{2}}
\right)^{2} \right]^{2}
$$
$$
\hspace{3.5cm} +\,H_{T}^{2}\,\left(\frac{\phi_{T}}{\phi_{F}}
\right)^{4}\left(\frac{S_{1}}{\varepsilon\phi_{F}^{2}}
\right)^{2}.
$$
We choose the positive root in Eq.~(\ref{r}), since with this root
we could get the appropriate Einstein's General Relativity limit
in which $\varepsilon\phi^{2}=\varepsilon\phi_0^{2}=Cte.\equiv
1/8\pi =1/\kappa^{2}$ with $\phi_{T} = \phi_{F}=\phi_0$. In this
limit the curvature radius  of the bubble wall becomes
$$
R\longrightarrow\,R_{GR}=\frac{S_{1}\kappa^{2}}{([H_{F}^{2}-H_{T}^{2}
+(S_{1}\kappa^{2}/2)^{2}]^{2}+H_{T}^{2}\kappa^{4}S_{1}^{2})^{1/2}}.
$$

A dimensionless quantity $\Delta\,s$, which represents the
strength of the wall tension in the thin-wall approximation, is
given in the Einstein's GR theory in ref.~\cite{re8}, which in our
case can be represented by
 \be \ds
\Delta\,s=\,\frac{S_{1}\,R}{2\,\varepsilon\,\phi_{T}^{2}}\,<\,1.
 \en

By numerically solving  the field equation associated with the JBD
field $\phi$, Eq.~(\ref{ec4}), we obtain for $\omega=500$ the
following values $\phi_{T}=7.645$ and $\phi_{F}=6.023$. With these
values we find that $\Delta\,s=0,436$. Analogously, for
$\omega=2000$, we obtain  $\phi_{T}=17.247$ and $
\phi_{F}=15.276$, and thus we get $\Delta\,s=0,429$. In the second
model, it is found that, for $\omega=500$, $\phi_{T}=7.800$ and
$\phi_{F}=5.220$, which gives $\Delta\,s=0,506$, and for $
\omega=2000$, $\phi_{T}=17.310$ and $\phi_{F}=14.817$ we find
$\Delta\,s=0,478 $. We should  note here that, as long as we
decrease the value of the $\omega$ parameter the strength of the
wall tension $\Delta\,s$ increases. We could see this from the
fact that in the Einstein's GR theory, $\triangle s$ becomes given
by $\triangle s_{GR}
 = 4 \pi R S_{1}$~\cite{MSTTYY}, which turns out to be smaller than the corresponding
 expression in the JBD theory, since the quantity $(\varepsilon\,\phi_{T}^{2}){-1}$
 increases. In the first of the particular cases described above, we get that,
 when $\triangle s $ is compared with  the corresponding value in the Einstein's GR theory,
 we find that $\triangle s -\triangle s_{GR}\simeq$ 0.012, for $\omega = 500$,
 and for $\omega$ = 2000 we get $\sim$ 0.005. In the second model this difference
 becomes of the order of $\sim 0.06$, for $\omega = 500$  and  $\sim 0.032$,
 for $\omega = 2000$.

%%%%%%%%%%%%%%%%%%%%%%%%%%%%%%%%%%%%%%%%%%
%                                        %
% ECS. DE CAMPO  LORENTZIANAS            %
%                                        %
%%%%%%%%%%%%%%%%%%%%%%%%%%%%%%%%%%%%%%%%%%

\section{\label{sec:level3}Inflation after tunnelling}

After the tunnel has occurred, we should both make an analytical
continuation to the Lorentzian spacetime  and see what is the time
evolution of the scalar fields $\phi(t)$ and $\sigma(t)$, and of
the scale factor $a(t)$. The field equations of motion for the
fields $\sigma$, $\phi$ and $a$ are given by \be \ds
\ddot{\sigma}\,=\,-3\,\frac{\dot{a}}{a}\,\dot{\sigma}\,-\,\frac{dV}{d\sigma}\,\,\label{ec7},
\en \be \ds
\ddot{\phi}=-3\frac{\dot{a}}{a}\dot{\phi}-\frac{\dot{\phi}^{2}}{\phi}
-\frac{1}{1+6\varepsilon}\left[\frac{\dot{\sigma}^{2}}{\phi}-\frac{4}{\phi}V(\sigma)
\right]\,\, \label{ec8},
 \en
and \be \ds
\ddot{a}=2\frac{\dot{a}\dot{\phi}}{a\phi}-\frac{a}{3\varepsilon\phi^{2}}\left[
\dot{\phi}^{2}+\frac{1+3\varepsilon}{1+6\varepsilon}\dot{\sigma}^{2}-
\frac{1-6\varepsilon}{1+6\varepsilon}V(\sigma)\right] ,\label{ec9} \en where the
dots now denote derivatives with respect to the cosmological time.

\begin{figure}[ht]
\includegraphics[width=3.0in,angle=0,clip=true]{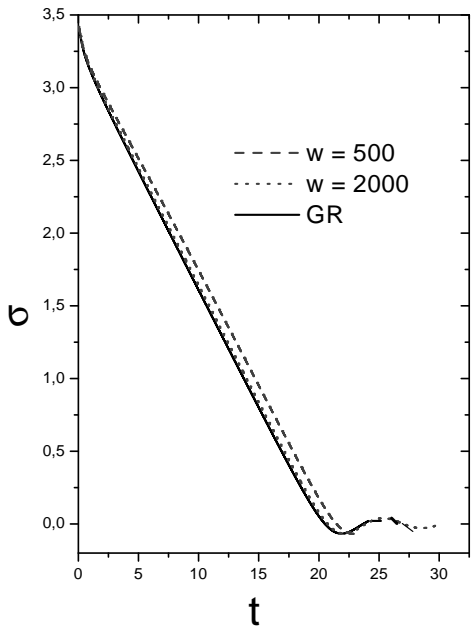}
\caption{This plot shows the inflaton field $\sigma$ as a function
of the cosmological time $t$ during the slow-rolls down open
inflationary universe, either in the Einstein's GR and JBD
theories ($\omega$=500 and $\omega$=2000). In these graphs we have
assumed the constant $m$ to be 1.} \label{fig7}
\end{figure}

In order to numerically solve this set of Equations we use the
following boundary conditions  $\dot{\sigma}(0)= \dot{\phi}(0)=0$,
$a(0) =0$ and $\dot{a}(0)=1$. In our first model the solutions
$\sigma(t)$ are shown in Fig.~\ref{fig7} for some different values
of the $\omega$ parameter. In the same situation, we have studied
the evolution of the JBD field, $\phi$. We have found that this
field monotonically increases to some constant value, which is
closer to that determined by the actual value of the Planck mass,
(recall that $\varepsilon\phi_0^{2}=1/8\pi$), just when the
inflaton scalar field $\sigma$ begins to oscillate near the
minimum of the effective potential, located at $\sigma \approx 0$.
We have also found that for the range $50\leq \omega \leq 2000$,
the universe could inflate more than the $60$ e-folding, which we
find in Einstein's theory of gravity. However, for a sufficiently
small value of this parameter, say $\omega \approx 5$ or so, the
e-folding obtained after tunneling has occurred, is not enough to
solve the cosmological puzzles, such as flatness, horizon, etc.
Therefore, we have found that our models are quite sensitive to
the value we assign to the $\omega$ parameter.

\section{\label{sec:level4}Scalar perturbation spectra}

Even though the study of scalar density perturbations in open
universes is quite complicated~\cite{re8}, it is interesting to
give an estimation of the standard quantum scalar field
fluctuations inside the bubble for our scenarios. The
corresponding density perturbation in the JBD theory
becomes~\cite{re1} \be
\frac{\delta\,\rho}{\rho}\,\approx\,Cte\,H^{2}\left[
\left(\varepsilon\phi^{2}8\pi\right)^{\frac{3}{2}}\,\frac{1}{|\dot{\sigma}|}\,
+\frac{(1-\varepsilon\phi^{2}8\pi)}{2|\dot{\phi}|}\gamma\phi\right]
,\label{ec10}
 \en
where $\ds \gamma\,=\,\frac{1}{\sqrt{\omega+3/2}}$ and $Cte.
\approx 3/5\pi$. The latter equation coincides with its analogous
equation in Einstein's theory, when the substitution
$\varepsilon\phi^{2}=1/8\pi$ is made. The reason why this
expression is approximated is because it is expected that other
contributions to the exact expression exist~\cite{re9}. However,
as was observed by Linde~\cite{re8}, we may use the above
expression for $N > 3$, as a correct result.

\begin{figure}[ht]
\includegraphics[width=3.0in,angle=0,clip=true]{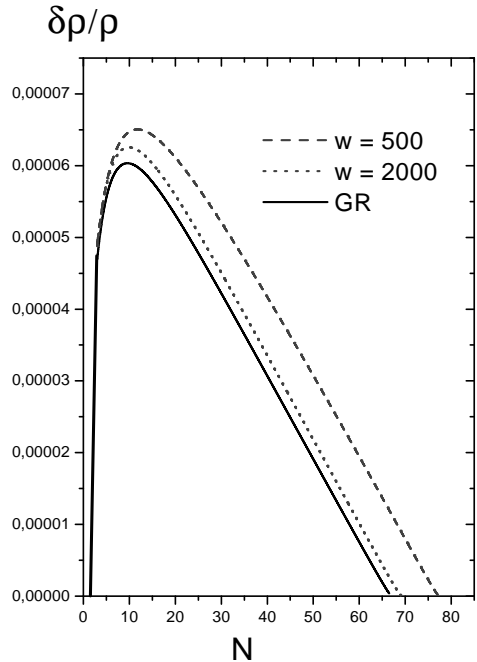}
\caption{Scalar density perturbations for our first model produced
inside the bubble N e-folds after the open universe creation. We
have plotted $\delta\rho$/$\rho$ for $\omega = 500$ and $\omega =
2000$. These plots are compared with those obtained by using the
Einstein's GR theory, where
$\delta\rho$/$\rho\approx\,CteH^{2}$/$|\dot{\sigma}|$}
\label{fig6}
\end{figure}

Figure~\ref{fig6} shows the magnitude of the scalar perturbations
$\delta\rho/\rho$ for our first model as a function of the $N$
e-folds of inflation for two different values of the $\omega$
parameter, after the open universe was formed. Even though the
shape of the graph is similar to the Einstein's GR case,
$\delta\rho/\rho$ has a maximum at small $N \sim O(12)$. Its
maximum value, however, increases a little bit, when we decrease
the $\omega$ parameter value. Similarly, the values of $N$
e-folds, where $\delta\rho/\rho$ vanishes, increase when $\omega$
decrease. We should mention that there is a relation between the
value of the scalar perturbation and the $N$ e-folds of inflation.
For $N \sim 10$, where $\delta\rho/\rho$ gets it maximum value, it
is found that the scale where the scalar perturbation is measured
corresponds to the  $10^{24} cm$ scale. However, for $N \sim 15$
it decreases to $10^{22} cm$, and for $N \gg 50$ this practically
comes to zero. We could show that something similar happens in the
second model considered. There, the corresponding values of $N$
e-folds were smaller.

\begin{figure}[ht]
\includegraphics[width=3.0in,angle=0,clip=true]{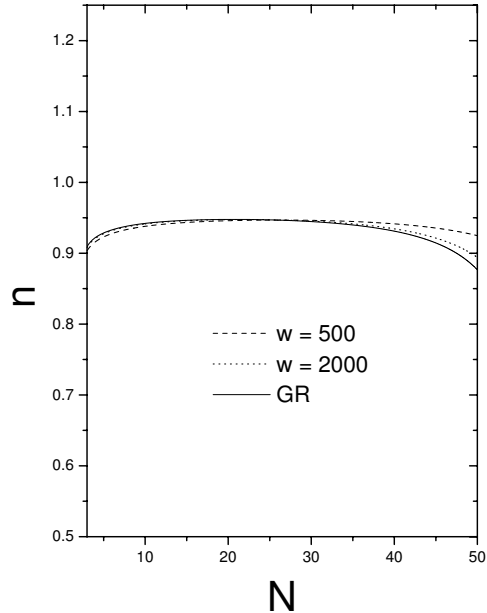}
\caption{The spectral index $n$ as a function of the  $N$ e-folds
parameter for two different values of the JBD parameter $\omega$.
The values that we have taken are $\omega = 500$ and $\omega =
2000$. These plots are compared with that obtained by using
Einstein's GR theory.} \label{nN}
\end{figure}

One interesting parameter to consider is the so-called spectral
index $n$, which is related to the power spectrum of density
perturbations $P^{1/2}_{\cal R}(k)$. For modes with wavelength
much larger than the horizon ($k \ll a H$), the spectral index $n$
is an exact power law, expressed by $P^{1/2}_{\cal R}(k) \propto
k^{n-1}$, where $k$ is the comoving wave number. In the slow roll
limit, where $\ddot{\sigma} \simeq 0$ and the first two
derivatives of the effective potential are small relative to its
magnitude, i.e., $V' $, $V''  \ll V $, with $V' = d V /d\sigma$,
it is found that the spectral index $n$ is given by
$$ n = 1 - 4 \epsilon + 2 \eta,
$$
where the parameters $\epsilon$ and $\eta$, the so-called slow
roll parameters are given by~\cite{KoVa}
$$ \ds \epsilon \simeq \frac{m_{Mp}^2}{16 \pi}\left(
\frac{V' }{V }\right)^2 \ll 1
$$
and
$$ \ds \eta \simeq \frac{m_{Mp}^2}{8 \pi}\left[
\frac{V'' }{V } - \frac{1}{2}\left( \frac{V'}{V}\right)^2 \right]
\ll 1.
$$
Figure~\ref{nN} shows the spectral index parameter $n$ as a
function of the e-folds parameter $N$ for two different values of
the JBD parameter $\omega$. With the aim of comparing, we have
also included here the spectral index in Einstein's GR theory.
Note that the parameter $n$ gets values which are, on average,
smaller than that found in the Einstein theory.

Certainly, apart from the scalar perturbations, tensor
perturbation also exist. These perturbations are usually
associated with perturbations of the bubble wall~\cite{re6}.
Specifically, in the Einstein's GR theory it is known that the
fluctuations of the bubble wall contribute to the low frequency
spectrum of tensor perturbations, which can dominate over the
scalar perturbations~\cite{YaSaTa,Ga,Ga-Be}. Here, we expect
something similar to occur in our models, except at low enough JBD
parameter, where the Einstein's GR and the JBD theories can be
distinguished one from the other. However, due to present bound of
the observational limits from the solar system measurements for
the $\omega$ parameter~\cite{Wi}, we expect these contributions to
become tiny corrections of that obtained in the Einstein's GR
theory. Certainly, this latter point deserves further
investigation that we hope to carry out in a near future.

\section{conclusion}

Since we still we do not know the exact value of the $\Omega$
parameter, it is convenient to count on an inflationary universe
model in which $\Omega < 1$. In this sense, we could have
single-bubble open inflationary universe models, which may be
consistent with a natural scenario for understanding the large
scale homogeneity and isotropy structure. However, open
inflationary models have a more complicated primordial spectrum
than that obtained in flat universes, where  extra discrete modes
and possibly large tensor anisotropies spectrum could be found,
especially those related to supercurvature modes, which are
particular to open inflationary universes. Forthcoming
astronomical measurements will determine if this extra terms are
present in the scalar spectrum.

In this paper we have studied one-field open universe models in
which the gravitational effects are described by a JBD theory. In
this theory the fundamental quantity is the JBD field $\phi$, from
which, after that universe enters the Lorentzian era, it can
numerically be shown that it monotonically increases from an
initial value to the present value of the Planck mass obtained at
the end of inflation. We have studied solutions to two effective
potentials in which the CDL instantons exist. The existence of
these instantons is shown because the inequality $|V^{''}|>H^{2}$
is satisfied, and thus, slow-roll inflationary universes are
realized for different values of the JBD $\omega$ parameter.

For the two models considered, $V^{''}$ remains greater than
$H^{2}$ during the first e-folds of inflation. In the thin-wall
limit we have also found an increase in the strength of the wall
tension, $\triangle s$, when compared with their analogous results
obtained in the Einstein's GR theory.

Since in $\delta\rho/\rho$ graphs the maximum present a small
displacement in the JBD theory, when compared with that obtained
in the Einstein's GR theory, this would change the constraint on
the value of the parameter $m$ that appears in the scalar
potentials. In this way, we have shown that one-field open
inflationary universe models can be realized in the JBD theory.

\begin{acknowledgments}
S.d.C. was supported from COMISION NACIONAL DE CIENCIAS Y
TECNOLOGIA through FONDECYT Grant Nos. 1000305 and 1010485. Also,
it was partially supported by UCV  Grant No. 123.752. R.H. is
supported the MECESUP FSM 9901 project.
\end{acknowledgments}

\end{document}